\documentstyle[hx37_latex]{article}
\input{psfig}
\begin{document}

\small
\vspace*{-3.0cm}
\noindent
\hspace*{3.4cm} To be published in ``HST and the High Redshift Universe''\\
\hspace*{7.3cm} Cambridge, UK, 1--5 July 1996 
\vspace*{0.5cm}
\normalsize

\title{
CONSEQUENCES OF A NON-DETECTION OF FORMING GALAXIES BY AN INFRARED SURVEY}

\author{F. MANNUCCI}
\address{CAISMI--CNR, Largo E.Fermi 5, 50125 Firenze, Italia}
\author{D. THOMPSON, S.V.W. BECKWITH}
\address{Max-Planck-Institut f\"ur Astronomie, K\"oningstuhl 17, 
		 69117 Heidelberg, Germany}
\maketitle

\abstracts{
We present the results of a narrowband search for forming
galaxies at high redshifts. Given the coverage of 276 square minutes of arc,
this is the most extended search for extragalactic emission line objects
at near infrared wavelengths.
Despite of flux limits down to 1.4$\times10^{-16}$\,erg\,cm$^{-2}$\,s$^{-1}$
and of a total comoving volume surveyed of $1.4\times10^5$\,Mpc$^3$
(for H$_0 = 50$\,km\,s$^{-1}$\,Mpc$^{-1}$, $\Omega_0 = 1$), no
such population was detected.  
We show how this null detection can be used to derive upper limits to 
a) the comoving volume density of this population and 
b) to the metal production density at redshift between 1 and 4.
}

\section{Introduction}

One of the most important issues of galaxy formation is understanding
when the majority of the stars of early-type galaxies formed
because identification of the first generations of stars and measurement 
of their ages are of great importance for models of the universe and 
galaxy formation.  

A genuine population of forming galaxies has been discovered by 
Steidel et al. (1996, ApJ 462, L17).
Our complementary approach consists in looking for redshifted optical 
emission lines in the near-IR.  The 
main advantage of this approach with respect to the optical searches
is that these lines are emitted at much 
longer rest-frame wavelengths and, therefore, suffer 
considerably less from extinction by dust. 
As an example, while the high-redshift galaxies detected by 
Steidel et al. (1996) appear to have fairly significant star formation rates,
they generally have weak or absent Ly$\alpha$ emission.

\section{Observations and data reduction}

We obtained deep images of selected fields through narrow and broad 
band filters and looked for objects which are relatively brighter in 
the narrow filter, thus indicating a substantial flux in an emission line.  
We chose the fields to contain objects with known 
redshifts which put optical emission lines in the 
passbands of the narrow filters.  If there is any tendency toward 
clustering, these objects will pinpoint regions of overdensity.

A total of 30 image pairs were 
obtained at the Calar Alto 3.5m and ESO/MPI 2.2m telescopes
with exposure times of 1-2 hours through the narrowband 
filter and 15-30 minutes through the broadband filter.  
Plots of the (broad$-$narrow) color vs. narrowband magnitude were 
constructed for each of the 30 image pairs.  Objects with relatively 
strong emission lines stand away from the locus of the remaining objects.  
There was only a single strong candidate identified in the survey data,
but this is a complex object and will be discussed in Beckwith et al.
(1997, in preparation).

This survey searches about two orders of magnitude more 
volume in the universe than similar infrared surveys previously completed.  
More details about observations, data reduction and results can be found
in Thompson et al. (1996, AJ, in press).

\section{Limits on the comoving volume density of forming galaxies}

If the detected object is not a forming galaxy, we can put upper limits on the
comoving volume density of such objects versus their star formation rates.
To calculate the volume sampled by the survey, 
we include the five strong lines: H$\alpha$, H$\beta$, [OIII]$\lambda5007$,
[OII]$\lambda3727$, and Ly$\alpha$.  Each of these would appear in the 
narrowband filters at different redshifts.  
The survey limits plotted in figure 1 assume no obscuration by dust, 
and therefore represent lower limits to the true star formation rate.  
The arrow in fig. 1 shows the effect that an extinction of 
$E_{(B-V)} = 0.3$ would have on the plotted survey limits.

These upper limits are compared to the expected density and luminosity 
of the forming galaxy population,
taking into account various star formation histories and mass evolution.  
The model are described in Mannucci and Beckwith (1995, ApJ 442, 569)
and estimate the density of young galaxies which are
necessary to produce the local population of elliptical galaxies.  
We consider three classes of models:
the first and simplest model (labelled {\em constant} in fig. 1)
assumes no evolution in the mass function of elliptical galaxies and a 
constant star formation rate for each galaxy during a certain 
period of time. 
The boundaries for each model in fig. 1 correspond to the maximum and 
minimum reasonable span of time, i.e.,
to the minimum and maximum expected brightness, respectively.  
The other two models correspond to mild evolution of the mass function
of the elliptical galaxies: in the {\em burst} model there are
more massive but less numerous objects, and vice versa for the
{\em hierarchical} model.

\begin{figure}
\centerline{\psfig{figure=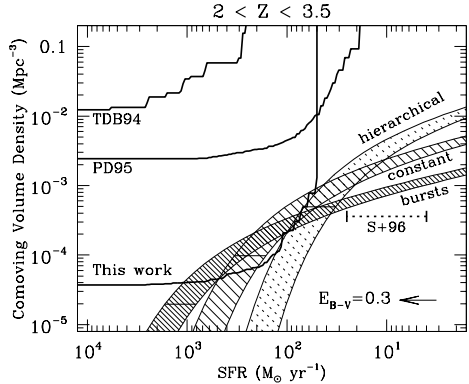,height=5cm}}
\caption{
Upper limits on the volume density of forming galaxies in the redshift 
range $2.0 < z < 3.5$.  
The {\em constant, bursts} and {\em hierarchical} models are discussed 
in the text.
The dashed line marks the comoving density and range of SFR for the 
population of star-forming galaxies detected by Steidel et al. (1996)
at $3.0<z<3.5$. The three thick lines are 
the upper limits to the PG volume density from three surveys (this work, 
TDB94: Thompson et al. (1994, AJ 107, 1); 
PD95: Pahre and Djorgovski (1995, ApJ 449, L1)),
where the regions to the upper left of these curves are excluded 
by the surveys.  The arrow shows the effect on these limits of an
extinction of $E(B-V)=0.3$.
}
\end{figure}

Figure 1 shows survey limits and model expectations in the 
redshift bin $2.0<z<3.5$.
In this redshift range, sampled mainly by the H$\alpha$ line in the K
band and by the H$\beta$ and [OIII] lines in the H and K bands,
objects with unobscured star formation rates equal to 
100\,M$_\odot$\,yr$^{-1}$ would be readily detected.  
Figure 1 shows that the data sample enough volume to 
exclude the {\em constant} and {\em burst} models, although the limits 
only partially overlap the expectations for {\em hierarchical} 
models.  
This means that unobscured young galaxies can be present in this redshift 
range only if either they are small systems, or have low surface brightness, 
or are less efficient in emitting lines.  Alternatively, the majority of 
galaxy formation could have occurred at higher redshifts.

\begin{figure}
\psfig{figure=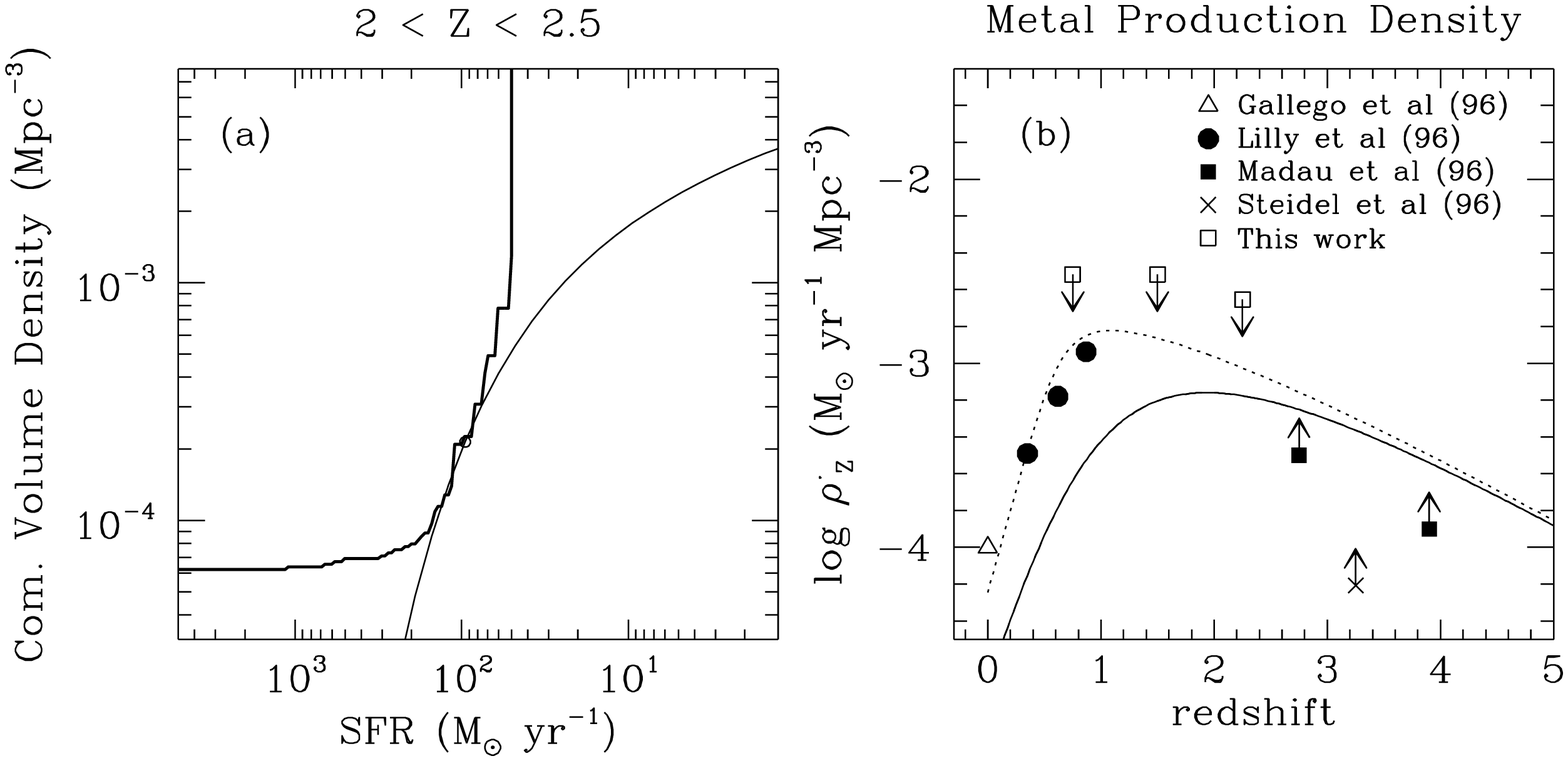,width=12cm}
\caption{
(a): measure of the maximum SFR compatible with our data: the SFR
function has $\phi^*=0.0087 h^3 Mpc^-3$ (1/3 of the value
for field galaxies given by Ellis et al., 1996, MN 280, 235) and 
SFR$^*=95$ M$_{\odot}$/yr, derived by matching to our survey limits.
(b) Upper limits to the metal production density vs. redshift from 
our survey compared with the models by Pei and Fall (1995, ApJ 454, 69)
(solid line: closed and outflow models; dotted line: inflow model)
and with other measurements and lower limits
(see Madau et al, 1996, for references).
}
\end{figure}

\section {The comoving metal production density}

The same data can be used to compute upper limits to the
density of metal production to be compared with other measurements and
lower limits.  The procedure is shown in fig 2.
We assume that both luminosity and SFR are proportional to the mass of a
galaxy, and that the mass function for the elliptical galaxies 
does not evolve strongly with redshift (i.e., we can neglect 
effects like inflow, outflow or merging).
This supplies us with the $\alpha$ and $\phi^*$ values of a Schechter 
``SFR function'' 
(number of galaxies with a given SFR vs. SFR) at any redshift.
We compute the maximum value of the third parameter SFR$^*$,
i.e., the maximum SFR of an $L^*$ galaxy giving no detection in our
survey at a given redshift (changing this value corresponds
to shifting these SFR function horizontally in fig. 2a)
By integrating the resulting SFR function
we compute the upper limit to the total SFR density and, using the
recipe in Madau et al. (1996, ApJ, in press), the metal production density.
We apply this procedure for those redshift ranges in which our survey
is sensitive, obtaining the upper limits in fig. 2b.

Even if these points are subject to some uncertainties (discussed in
Mannucci et al., 1997, in preparation) due to the various
assumptions, it seems that they can put intertesting constraints on the 
models of galaxy formation

\section{Conclusions}

Young galaxies with emission lines at infrared wavelengths are rare at 
the level that can be reached with the current generation of detectors. 
The failure to detect many young galaxies makes it unlikely that most 
galaxies had star formation histories with continuous formation 
starting at any redshift and continuing to about $z \sim 2$.  Hierarchical 
formation, in which galaxies were assembled from many pieces over a 
long interval are consistent with the results. 

It is also possible that physical conditions not included in the models 
could reduce the observable line flux and
weaken the conclusions derived.  Two examples:
a) dust along the lines of sight;
b) young galaxies might be very extended, making the surface 
brightness too low to see in this survey. 

\end{document}